# Local Charged States in $La_{0.89}Sr_{0.11}MnO_3$ Single Crystals

R. F. Mamin[1,2] *, I. K. Bdikin[1], A. L. Kholkin[1]

[1] *Department of Ceramics and Glass Engineering & CICECO, University of Aveiro, 3810-193 Aveiro, Portugal*

[2] *Zavoisky Physical-Technical Institute of RAS, 420029 Kazan, Russia*

We report the direct evidence of the possibility of charge segregation in manganites. The high contrast of electric-field-induced local charged states is observed during more than 100 hours at room temperature. These induced states display a piezoelectric response and polar properties. This effect may be an example of a new kind of polar ordering in solid state.

The complicated interplay among charge, spin and lattice degrees of freedom in manganites is believed to induce the unexpected magnetic and transport phenomena, such as the colossal magnetoresistance (CMR) [1–6], which is one of the most notable phenomenon in solid-state physics since the discovery of the high-temperature superconductors. While the tendency to the inherent phase and charge segregation is widely discussed for both - manganites [3–6] and high-temperature superconductors [7-10], clearly the microscopic origin of this extraordinary behavior is far from being understood and the question about the possibility and temperature region of these phenomena remains to be clarified [3–6, 12]. These inhomogeneities can appear even above the Curie temperature and the new scale of temperature $T^*$ should be introduced [5, 7]. In the manganites below $T^*$ the metallic clusters start to form and their percolation may lead to the long-range order as well as the CMR effect. In low-doped manganites the percolation is not complete and moreover, the characterictic temperature $T^*$ can be above the room temparute. In this letter we directly show that the tendency to the formation of the long-lived areas with the increased charge concentration appears in manganites already at *room* temperature.

The perovskite manganite $La_{1-x}Sr_xMnO_3$ crystals became the model object for such kind of investigations [5, 6]. It is known [15-19] that the $La_{0.89}Sr_{0.11}MnO_3$ (LSM-0.11) composition has a complex phase behavior and undergoes several transitions upon cooling. At $T_{RO} \approx 520$ K the structural transition from the rhombohedral (*R*) to the weakly distorted orthorhombic (*O*) phase occurs. At $T_{OO'} \approx 310$ K the structural transition from the weakly distorted orthorhombic (*O*) phase to the strongly distorted Jahn-Teller orthorhombic (*O'*) phase occurs. In this compound only the metallic nano-patches can be formed at $T^* \approx T_{OO'}$, which is above room temperature. This circumstance has allowed us to use scanning probe microscopy (SPM) techniques at *room* temperature for the induction and investigation of the local charged states.

Recently the colossal magnetocapacitive effect (CMC) was found in manganites [11–13] and other materials [14]. While the search for materials with the magnetoelectric effects has mainly focused on insulating materials [11, 14], we attempt to attain such effects at room temperature through the manipulation of intrinsic inhomogeneity [5, 12]. It is expected that important effects on electrical properties can be possible if the concentration of such carriers is locally varied in a controllable way. Therefore the attempts for inducing and characterizing these states were made by the SPM techniques in the electric-field-induced regime. We find that the electric-field-induced local polar state displays in LSM-0.11 at the *central symmetric* orthorhombic phase. We believe that this is one more evidence that the charge ordering occurs on nanoscale.

The SPM measurements were carried out on an LSM-0.11 single crystal with (001) and (100) orientations. The (001)-oriented sample has a cylindrical shape and thickness 1 mm. The (100)-oriented sample has a shape of a parallelepiped with the *c*-axis parallel to the long axis and the dimension of 5×3×1 mm$^3$. These crystals were characterized in earlier works [6, 12]. The measurements were performed in a commercial setup Multimode, NanoScope IIIA, DI. The local electrical properties of the samples are obtained with SPM techniques including Kelvin probe force microscopy (KFM) [20] and piezoresponse force microscopy (PFM) [21, 22]. In the KFM mode the bottom electrode was grounded. When *dc* voltage was applied and in the PFM mode the tip was grounded. The magnetic measurements were carried out on a Quantum Design magnetic properties measurement system in the temperature range of 80-380 K.

A topographic image of the scanning area is shown in Fig. 1a. It does not reveal any changes after poling with the electric field less than 40 V. The original state of the samples does not reveal the contrast of images in KFM and PFM modes. The induced states were obtained by different voltages applied during the few scanning lines when the tip covers the 200 nm in the horizontal direction (see Fig. 1) during 10 s, thus the induced area was $200\times1000$ nm$^2$. The electric-field-induced contrast in the KFM images in the (001) orientation after scanning in the central part of a scanning area under the electric field of ±10 V is shown in Fig. 1b, confirming a local spatial redistribution of the charges. Electric-field-induced state in (100) oriented LSM-0.11 crystals is measured also and the magnitudes of the effect in (001) and (100) oriented crystals are similar. However, it is seen that the contrasts from positive and negative voltages are not equivalent (see Figs 1b and 2). Figure 2 shows the change of the cross sections of KFM images with time measured directly after poling, 100 and 163 h later. After several days, both KFM and PFM signals from induced states are still present. From the conductivity measurements the Maxwell relaxation time of these compound is estimated as $\tau_M = 10^{-9}$ s. The induced polar charged state relaxes with characteristic time constant of about 100 h, which exceeds the Maxwell relaxation time by many orders of magnitude.

Figure 1c shows the image of the piezoresponse contrast of the electric-field-induced state. The PFM signal appears directly in the poling scanning spot. The intensity of the piezoresponse signal is increased with increasing the poling voltage. The contrasts from positive and negative voltages are not equivalent (see Fig.1c) and repeat the situation in the KFM image. These measurements were complemented with local piezoelectric hysteresis measurements (see Fig. 3.). The change of the PFM signal in the point as the voltage of poling electric field in LSM-0.11 is changed has a ferroelectric character: with saturation and hysteresis as shown in Fig. 3. It is similar to that in ordinary ferroelectric materials [23]. But the magnitudes of maximum piezoresponse signals for the first and second loops are different. This can occur because the first polar state is formed from the nonpolarized state of the sample, while at the second and next switching a polarized state is formed from the state with an opposite polarization. Thus the hysteresis loops for LSM-0.11 crystals differ from those in standard ferroelectrics by the presence of long time relaxation. Due to the long time of relaxation to the polar state the polarization is relatively weaker, but the polar ordering of some volume near the surface is induced as confirmed by the measurements of piezoelectric hysteresis loops.

In addition, the surface magnetic topography is scanned while being monitored for the influence of magnetic forces in magnetic force microscopy (MFM) mode. These magnetic forces are measured using the principle of the force gradient detection. MFM contrast of the electric-field-induced state is clearly observed as shown in Fig. 1d. These measurements were complemented with the investigation of the temperature behavior of the magnetic system. The *ac* magnetic susceptibility $c$(T) and magnetization M(T) are plotted in Fig. 4. The maximum in $c$(T) and the increase of M(T) suggest a magnetic transition in the vicinity of 165 K. In addition, $c$(T) shows the obvious anomaly at the structural phase transition in the vicinity of 310 K.

Our results reveal two very unusual phenomena: (i) the electric-field-induced charged state with the long relaxation time which by many times exceeds the Maxwell relaxation time for the charge inhomogeneities and (ii) the occurrence of a polar state at the *central symmetric* orthorhombic phase. The possible explanation could be the appearance of the phase and charge segregation occur at the temperatures below $T_{OO'}$ ($T_{OO'} \approx 310$ K). In this case the nano-patches with the increased charge concentration are formed. There is the magnetic order in these nano-patches [1-3, 24, 25]. The possibility of such phase segregation is widely discussed for manganites of various compounds [5, 25, 26]. One of explanations of the origin of the phase segregation is the double exchange interaction [25, 5]; other considerations are proposed as well [5, 26]. But even on the basis of only the experimental phase diagram it is possible to conclude that the areas with the large concentration of current carriers have the ferromagnetic order and simultaneously the metal type of conductivity. It is important to note that at the high strontium concentration (x > 0.17) the percolation between metallic areas is discussed to occur and it explains the metal - semiconductor transition. This scenario allows one to explain the CMR effect [5]. We believe that the observed phenomena could be determined by the occurrence of charge and magnetic inhomogeneities on nanoscale.

In the case of low Sr concentration the percolation does not occur and only the nano-patches of a metallic phase with the enlarged positive charge at the negatively charged matrix with a smaller concentration of current carriers are formed. As the magnetization in separate patches does not correlate, the magnetic phase in the total volume does not arise. The features of the magnetic characteristics found at the 310 K are probably the result of such evolution on nanoscale [27]. At applying of an external electrical field these positively charged nano-patches shift with respect to the negative insulated matrix and then they are fixed on new positions. The surface charge and the charge on the interphase boundary stabilize this new structure. As a result, the area with polarization is induced in the sample. The charges on the interphase boundary screen this polarization. The observation of the piezoresponse indicates the bulk nature of the induced states. The measurements of a local piezoelectric hysteresis confirm the reversible reorientation of local polarization.

The injection and concentration of the conductivity holes in the area in the vicinity of the tip under the negative voltage will promote the appearance of the polar state. This process is effective when we put on a negative constant voltage with inducing the polar state, because the conductivity elements are holes. Thus when the ground is on the tip under negative voltage the holes accumulate in the inducing area. This stimulates the increase of the volume concentration of the positively charged nano-patches. This in turn increases the magnitude of the effect. In this way the difference in the magnitude of the effect at applying of the negative and positive voltages becomes understandable.

The additional charge at the induced state can be stabilized due to the additional magnetic order. The areas, which have been induced at V < 0, should have a higher magnetic order. This supposition is to be clarified experimentally in future. On other hand these charged nano-patches are formed by the conductivity charges, and this process occurs on nanoscale. Therefore it is possible to expect that this evolution has no effect for the lattice symmetry on the scale of an elementary cell, and in X-ray experiments we will observe the central symmetric lattice symmetry (the orthorhombic in our case) as before.

We cannot reliably establish the presence of magnetic order from our MFM experiments because of the strong influence of the Coulomb interaction on these measurements. However the results of magnetic susceptibility measurements (see also [12, 28] for x=0.125) confirm that the magnetic evolution occurs in the system at $T_{OO'}$. The Jahn-Teller interaction can promote the discussed evolution [5, 26]. We assume that the very long relaxation time for the charged state after poling could be explained by the occurrence of a self-assembled intrinsic inhomogeneous state. The stability of such states is determined by the mutual interaction of the segregated charges with magnetic local order, which defines the actual long relaxation times of these states. The system in such states is in a relative potential energy minimum. Thus these results are the direct evidence that the states with localized charge on nanoscale can be stable on times much larger than the Maxwell relaxation time at room temperature for this compound.

In summary, the distinct contrast of the electric field induced charged states is observed in LSMO-0.11 single crystals confirming the possibility of charge segregation in manganites. The piezoelectric contrast is observed in these states pointing to the existence of a local polar state at some volume. The injection of the additional conductivity currents in the induced area promotes the occurrence of the polar charged states. The long relaxation time for the induced charged state might be explained by the existence of the intrinsic inhomogeneous states. All these results show that there is a tendency to the formation of the stable areas with the increased charge concentration. While the appearance of charged states seems obvious, the magnetic properties of these states remain to be clarified.

The authors wish to acknowledge M.I. Balbashov for providing the LSM-0.11 single crystals used in this study. R.F.M. thanks A. M. Bratkovsky, V. V. Kabanov and G. B. Teitel'baum for fruitful discussions. The work was supported by NOE "FAME" (NMP3-CT-2004–500159) and by the RFBR (Russia, grant No. 05-02-17182). R.F.M. acknowledges the financial support from NATO (grant CBP.NR.NREV 982014).

______________________________

*Electronic address: mamin@kfti.knc.ru

Figure captions:

Fig. 1. SPM images of the electric-field-induced contrast in a 001-oriented LSMO single crystal, left-hand line induced at $V = -10$ V and right-hand line – at $V = +10$ V: (a) topographic image before poling (it is the same after poling); (b) KFM image after poling; (c) PFM image after poling; (d) MFM image after poling.

Fig. 2. Relaxation of KFM images with time in a 001-oriented LSMO single crystal: cross sections of the images taken directly after poling, 100 and 163 h later.

Fig. 3. Local piezoresponse hysteresis loops: ABCD - the first and DEFG - second loops.

Fig 4. Temperature dependence of the *ac* magnetic susceptibility $c$ (*H*//*c*-axis) and the inverse *ac* magnetic susceptibility $c^{-1}$; the inset shows the temperature dependence of the magnetization at different fields (2T and 4T).

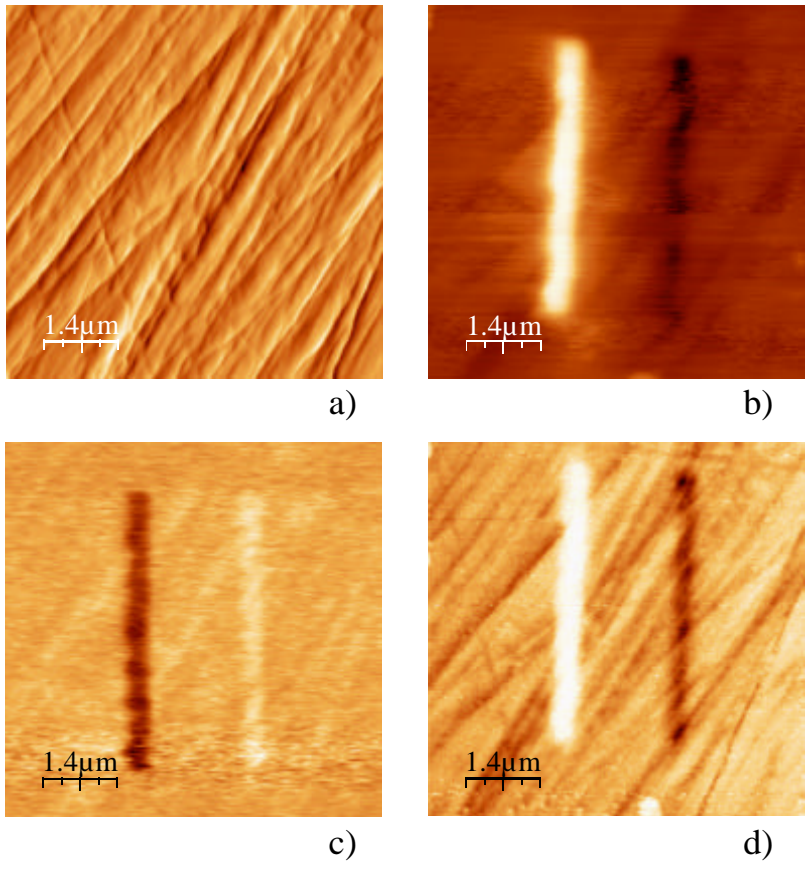

Fig. 1

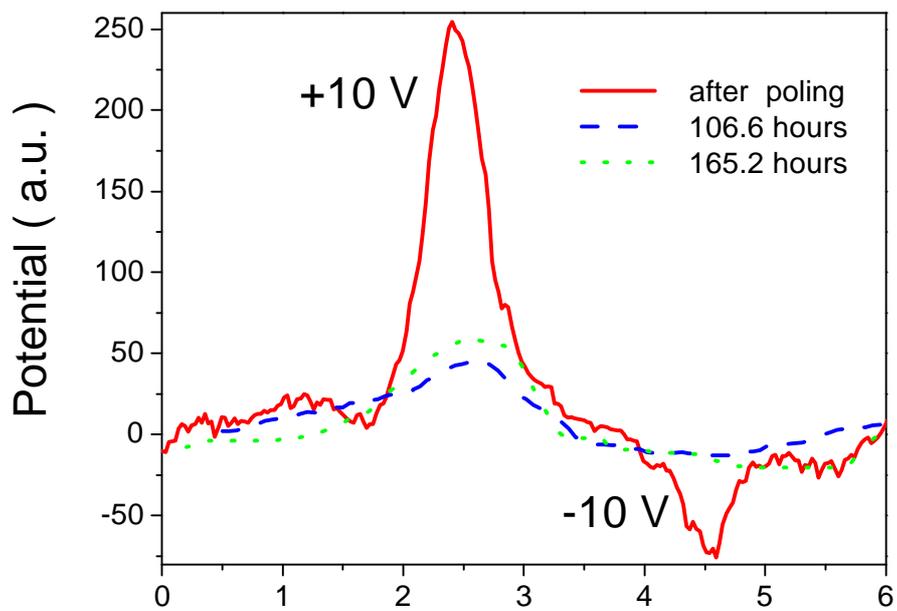

Fig. 2

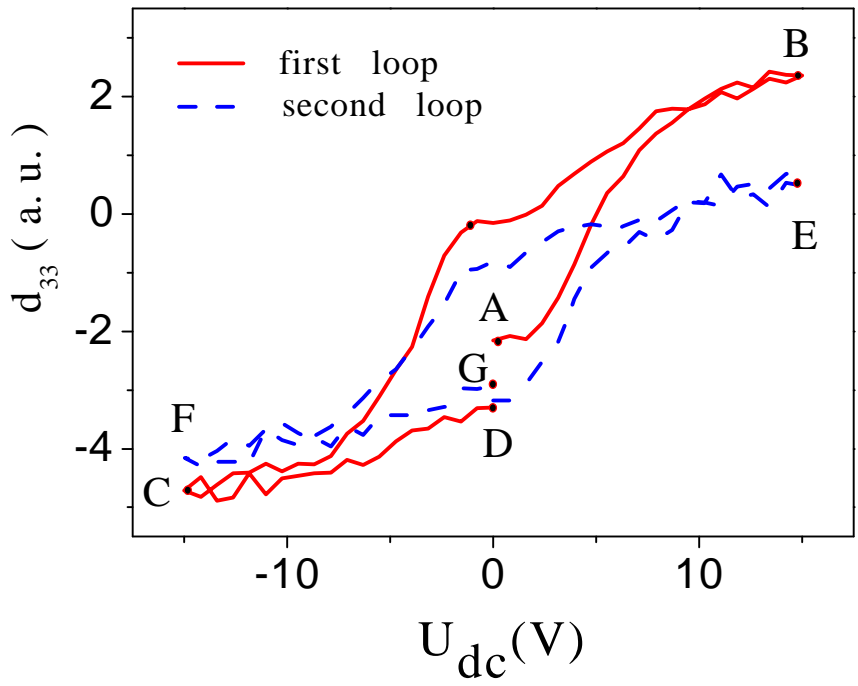

Fig. 3

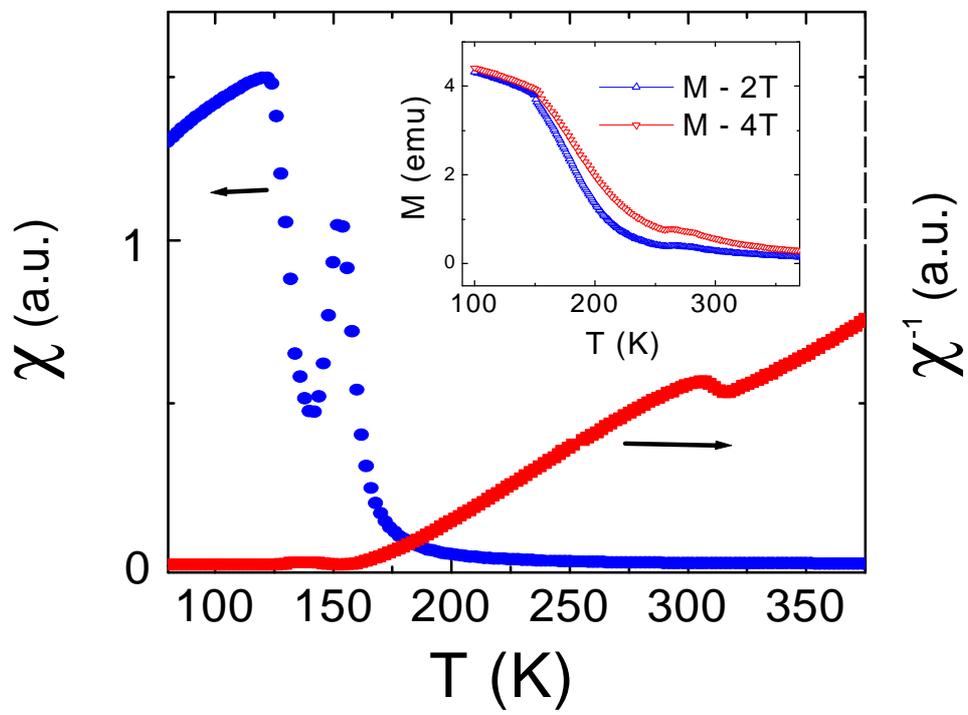

Fig. 4